\documentclass[8.5pt,twoside,twocolumn]{article}

\usepackage[super,sort&compress,comma]{natbib} 
\usepackage{mhchem}
\usepackage{times,mathptmx}
\usepackage{sectsty}
\usepackage{balance} 
\usepackage{subfigure}
\usepackage{graphicx}
\usepackage{color}
\usepackage{lastpage}
\usepackage[format=plain,justification=justified,singlelinecheck=true,font=small,labelfont=bf,labelsep=space]{caption} 
\usepackage{fancyhdr}
\usepackage[T1]{fontenc}
\usepackage[utf8]{inputenc}
\usepackage{hyperref}
\usepackage{amsmath}

\begin{document}

\makeatletter 
\def\subsubsection{\@startsection{subsubsection}{3}{10pt}{-1.25ex plus -1ex minus -.1ex}{0ex plus 0ex}{\normalsize\bf}} 
\def\paragraph{\@startsection{paragraph}{4}{10pt}{-1.25ex plus -1ex minus -.1ex}{0ex plus 0ex}{\normalsize\textit}} 
\renewcommand\@biblabel[1]{#1}            
\renewcommand\@makefntext[1]%
{\noindent\makebox[0pt][r]{\@thefnmark\,}#1}
\makeatother 
\renewcommand{\figurename}{\small{Fig.}~}
\sectionfont{\large}
\subsectionfont{\normalsize} 

\fancyfoot{}
\fancyhead{}
\renewcommand{\headrulewidth}{1pt} 
\renewcommand{\footrulewidth}{1pt}
\setlength{\arrayrulewidth}{1pt}
\setlength{\columnsep}{6.5mm}
\setlength\bibsep{1pt}


\twocolumn[
  \begin{@twocolumnfalse}
\noindent\LARGE{\textbf{Elasto-capillary meniscus: Pulling out a soft strip \\ sticking to a liquid surface}}
\vspace{0.6cm}

\noindent\large{\textbf{Marco Rivetti$^\dag$
and
Arnaud Antkowiak$^\star$}}\vspace{0.5cm}

\noindent\large{\textit{Institut Jean Le Rond 
d'Alembert, Unit\'e Mixte de Recherche 7190, Universit\'e Pierre et Marie Curie and Centre National de la 
Recherche Scientifique, 4 Place Jussieu, F-75005 Paris, France.}}\vspace{0.5cm}

\noindent \normalsize{A liquid surface touching a solid usually deforms in a near-wall meniscus region. In this work, we replace part of the free surface with a soft polymer and examine the shape of this `elasto-capillary meniscus', result of the interplay between elasticity, capillarity and hydrostatic pressure. 
We focus particularly on the extraction threshold for the soft object. Indeed, we demonstrate both experimentally and theoretically the existence of a limit height of liquid tenable before breakdown of the compound, and extraction of the object. Such an extraction force is known since Laplace and Gay-Lussac, but only in the context of rigid floating objects. We revisit this classical problem by adding the elastic ingredient and predict the extraction force in terms of the strip elastic properties. It is finally shown that the critical force can be increased with elasticity, as is commonplace in adhesion phenomena.
%
}
\vspace{0.5cm}
 \end{@twocolumnfalse}
  ]
 

\section{Introduction}

\footnotetext{\dag~Present address: Surface du Verre et Interfaces, UMR 125, CNRS \& Saint Gobain, 93303 Aubervilliers, France. Email: marco.rivetti@saint-gobain.com}

\footnotetext{$\star$~arnaud.antkowiak@upmc.fr}







Because interfaces are sticky, it is more difficult to pull out an object 
resting on a liquid surface than to simply lift it. This is a common illustration of capillary adhesion. The excess force actually corresponds to the weight of the liquid column drawn behind the pulled object, but relies intimately on surface tension as the liquid column is sculpted by capillarity into a \textit{meniscus}. 
Laplace in 1805 was the first to investigate this problem in his founding monograph on capillarity \citep{Laplace1805}. The theory designed by Laplace predicted a maximal height for the liquid meniscus above which no more solution could be found, hence a maximal force needed to extract an object. On Laplace's request Gay-Lussac undertook careful experiments on the traction of glass disks floating on water and obtained a beautiful agreement with this first ever theory of capillarity. 

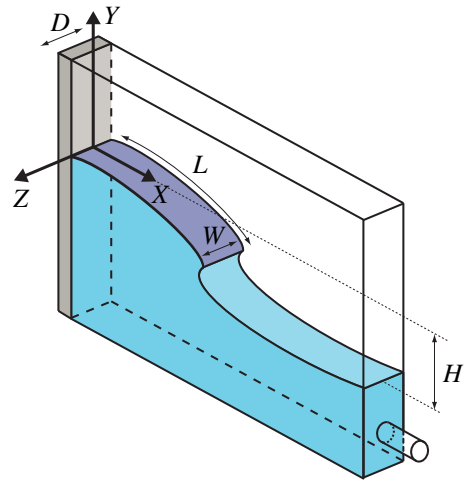
\begin{figure}
\begin{center}
\setlength{\unitlength}{.9cm}
\begin{picture}(7,7.1)
\put(0,0){\includegraphics[height=6.3cm]{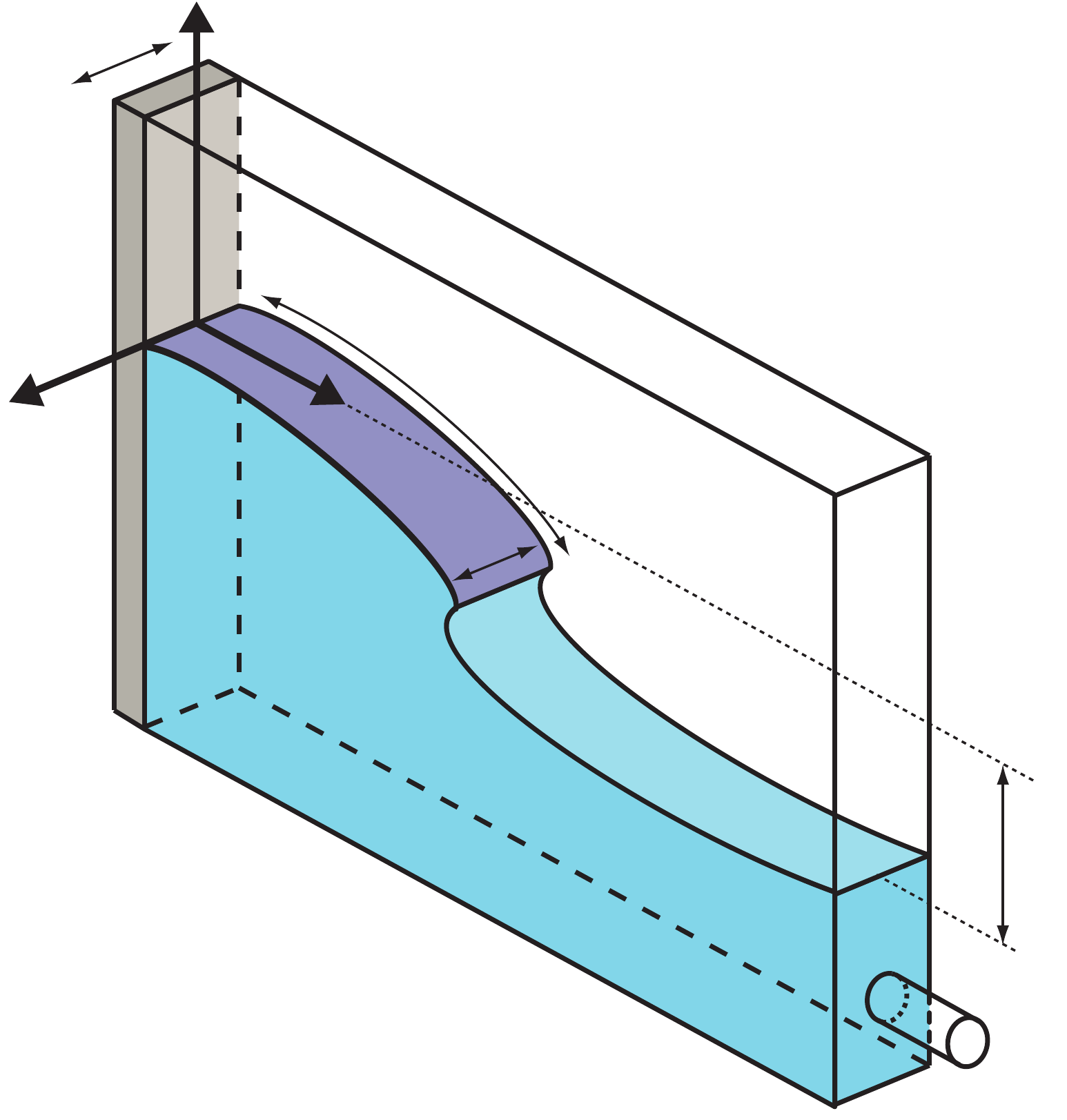}}
\put(2.7,4.6){$L$}
\put(2.1,4.15){$X$}
\put(1.4,6.8){$Y$}
\put(0.05,4.1){$Z$}
\put(0.6,6.7){$D$}
\put(6.4,1.5){$H$}
\put(2.85,3.55){$W$}
\label{fig:experiment_sketch}
\end{picture}
\end{center}
\caption{\textbf{Elastocapillary meniscus experiment.} A thin polymer strip (violet), of length $L$ and width $W$, is clamped inside a narrow rigid box partially filled with water (blue). Initially, the water level matches the embedding height. Upon withdrawal of the liquid, the soft strip and the free surface deform altogether into an elasto-capillary meniscus of height~$H$.}
\label{fig:experiment}
\end{figure}
%

Adhesion is not the only consequence of the existence of menisci. Collective behaviors such as the clustering of bubbles into rafts are also related to this capillary interface deformation. Such capillary attraction (or repulsion) has been known for long, and even used in the early 50's as a model for atomic interactions\citep{Nicolson1949}. In the case of anisotropic particles, clustering becomes directional and allows for self-assembly or precise alignment of particles\citep{Srinivasan2001, botto2012}. 

\begin{figure*}
    \centering
\includegraphics[width=18cm]{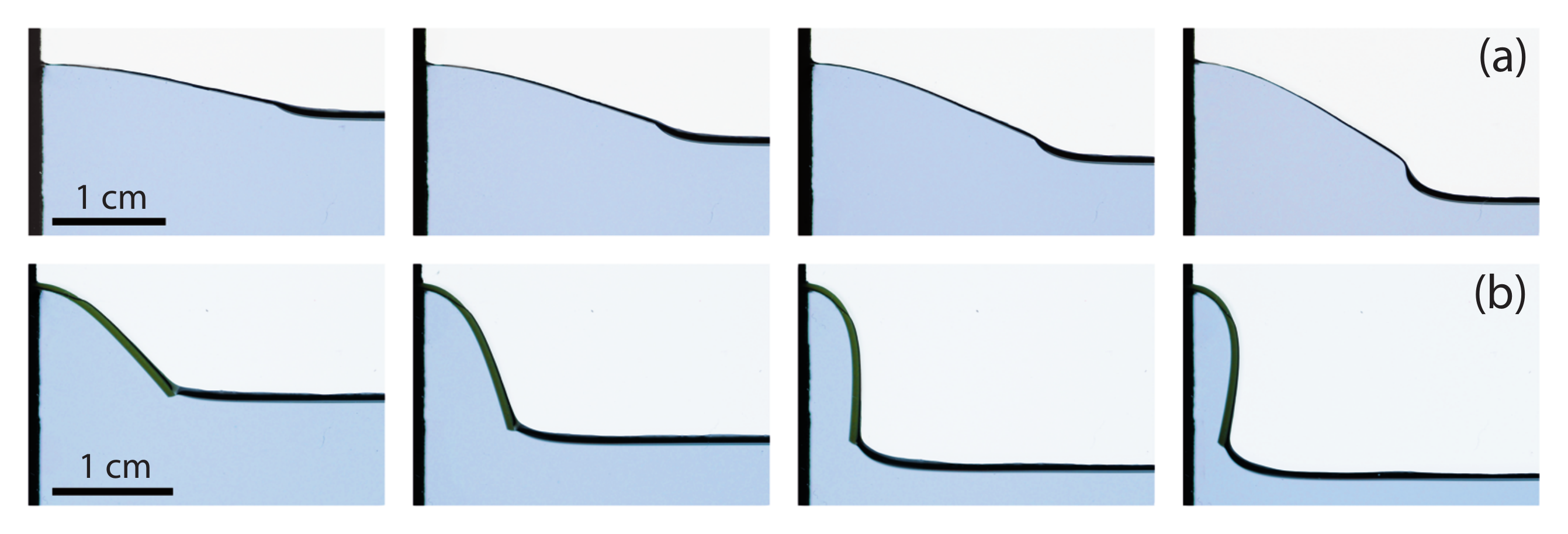}
\caption{\textbf{Typical elasto-capillary meniscus shapes.} (a) Equilibrium configurations for a Mylar strip floating on water. The length of the strip here is $L = 21$ mm. From left to right, the meniscus height is respectively $H= 5.0$, $7.2$, $8.9$ and $12.5$ mm.
     (b) Same snapshot sequence for a PVS strip resting on water. The length of the strip is $L = 15.8$ mm. From left to right, the meniscus height is respectively $H= 10.7$, $13.6$, $16.1$ and $16.6$ mm.
}
    \label{fig:experiment_snapshots}
\end{figure*}
%



In all these cases, there is an unavoidable elastic deformation imparted by the meniscus to the neighboring solid substrate, which may become non negligible for soft enough materials. 
Quite recently there has been a keen interest on the possibility to strongly deform an object with such elasto-capillary interactions. As a matter of fact, the variety of elasto-capillary driven shape distortion is surprisingly vast: clumping\citep{bico2004}, buckling\citep{neukirch2007}, 
wrinkling \citep{huang2007} or selectable packing\citep{Antkowiak2011},
as well as changing in wetting properties \citep{blow2010}
to quote a few examples, all being particularly relevant at small scales\citep{roman2010}.


Therefore, in contrast to Laplace and Gay-Lussac experiment, elastic structures resting on liquid interfaces may experience large deformations when  dipped in or withdrawn from the liquid, because of  interactions with hydrostatic pressure and surface tension. 
%
Nature offers an elegant example of such deformations with aquatic flowers
avoiding flooding of the pistil by folding their corolla as the level of water increases \citep{armstrong2002}. 
This mechanism recently inspired \citet{reis2010a} a smart technique to grab water from a bath using a passive \textit{`elasto-pipette'}.
%
A different but related problem is the case of a thin polymeric film draping over a point of contact on water\citep{Holmes2010}, showing the appearance of wrinkles and their transition towards localized folds. 
%
Elasticity is also the key to understand some surprising phenomena involving free surfaces, as for instance 
the difficulty to submerge hairs into a liquid by pushing them \citep{andreotti2011},
the water-repellent ability of striders' legs \citep{park2008_insects, Vella2008} 
or even the possibility for some insects and reptiles to walk on the water \citep{bush2006}. 
%
%
On this topic, \citet{burton2012} proposed recently a simple model to describe how elasticity modifies the buoyancy properties of a floating rod.

%

%
In the present study we propose a fully non linear description of the shape a soft thin object resting on (and deformed by) a liquid surface. 
We study particularly the quasi-static extraction of such an object from the liquid bath and we consider the deformed liquid-solid ensemble as an \textit{elasto-capillary meniscus} -- a generalization of the well known capillary meniscus including elasticity at the interface. 
%
%
In the following we will detail the different r\'egimes appearing in the problem, depending on the relative importance of elasticity, gravity and capillarity. Particular attention will be paid to the behavior of the contact line between liquid and solid, whose slippage can seriously influence the extraction process. %
We finally discuss the extraction process at the light of the particular r\'egime considered and show the possibility to increase the extraction force with a purely elastic mechanism, as is common in adhesion phenomena.

\begin{figure*}
    \centering
	   \includegraphics[width=16cm]{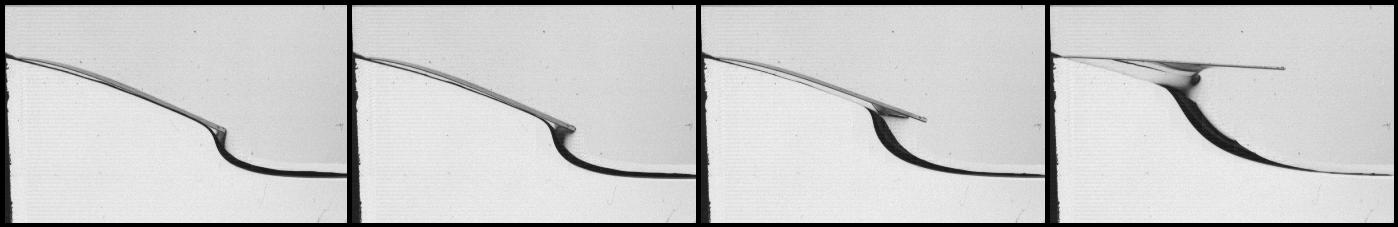}
   	 \caption{\textbf{Elastic strip extraction.} Sequence illustrating the failure of the elasto-capillary meniscus, associated with air invasion from the end of the strip and violent elastic relaxation. The time intervals between two consecutive images are 168 ms, 78 ms and 27 ms.}
	 \label{fig:collapse_classique}
\end{figure*}

\section{An elasto-capillary meniscus}
\label{sec:experiments}

In our experimental setup, sketched in Fig.~\ref{fig:experiment}, a narrow box with rigid glass walls is filled with water (density $\rho~=~10^3~\mathrm{kg/m^3}$ and surface tension $\gamma = 72~\mathrm{mN/m}$).
A thin and narrow strip,
of length $L$, width $W$ and thickness~$e$, is clamped at one end to the box, and is free at the other end. 
Strips employed in the experiments are made of polyethylene terephthalate (PET, bending stiffness per unit width $B~=~1.7~\cdot~10^{-4}~\mathrm{Nm}$) or polyvinyl siloxane (PVS, bending stiffness per unit width $B~=~1.8\cdot10^{-5}~\mathrm{Nm}$).
The lateral extent of the box is $D>W$, such that a small gap exits between the strip and the walls in order to avoid friction. 

The initial amount of water in the box is set to leave the liquid surface at the same level than that of the embedding in the wall, $Y=0$. The strip floats on water and is completely flat. 
Then, rather than to extract the strip, water is quasi-statically withdrawn from the box with a syringe (see Supplementary Information for a photograph of the setup), so that the liquid interface lies at distance $H$ from the embedding. The liquid-air interface pins to the free edge of the strip, causing its deflection.  Note that to ensure quasi-staticity, the level of the free surface is set to decrease at a rate lower than 2 mm/s.%


Focusing on the ($X,Y$) plane, we show Fig. \ref{fig:experiment_snapshots} several equilibrium configurations, obtained for one PET and one PVS strip, and for increasing values of the parameter $H$. Note that these pictures correspond to static configurations, as the withdrawal process has been stopped during the shot. The overall deformation of the liquid surface and the elastic strip results from a competition between capillarity, elasticity and gravity. It shall thus be referred to as an elasto-capillary meniscus in the following.

Looking at the shape of the free liquid surface, it is already quite clear that the total height of the compound can be much larger than that of a pure liquid meniscus. Note that there is no paradox however, as the height of the liquid portion still does not exceed  two times the gravity-capillary length $L_\mathrm{gc}~=~(\gamma / \rho g)^{1/2}$, where $g$ denotes the gravity. The elasticity of the strip is the factor allowing to reach such heights, and to lift large amount of liquid.

One can observe that the main difference between the sequence in panel \ref{fig:experiment_snapshots}(a) and \ref{fig:experiment_snapshots}(b) is related to the evolution of the contact angle between the strip and the liquid. Indeed, in panel \ref{fig:experiment_snapshots}(a) the angle decreases with respect to $H$, from a value near $180^\circ$ in the first image to approximatively $120^\circ$ in the last image. Conversely in panel \ref{fig:experiment_snapshots}(b) the contact angle increases with respect to $H$ up to approximatively $230^\circ$ in the last frame.
The variety of contact angle values is a consequence of the sharp edge at the end of the strip. This contact line pinning avoids air or water invasion in the region below or above the strip respectively, as already pointed out in \citet{reis2010a}, provided that the hysteresis of the contact angle is strong enough as we will discuss later. 

The height of the elasto-capillary meniscus can be increased up to a point where there is a sudden receding of the contact line associated with air invasion and a violent elastic relaxation, as depicted Fig.~\ref{fig:collapse_classique} (see also the corresponding movie in Supplementary Material). This is the extraction of the strip, which is not a continuous but a violent process occurring over a typical timescale of 100 ms. It is therefore a much faster phenomenon than the withdrawal process which takes tenth of seconds. Extraction occurs for a critical height depending a priori on all the parameters of the problem and will be examined in detail in section~\ref{sec:max_height}.
%
%
\section{Theoretical description}
\label{sec:theory}

In this section we present a 2D theoretical description of the shape of the elasto-capillary meniscus. On the one hand, the 2D approach for the elastic strip is justified by the fact that in all experiments $L \gg W$ and $L \gg e$, which allows for Euler-Bernoulli description of a beam. On the other hand, one can notice that the liquid interface is also deformed in the $Z$ direction, as suggested by the thick dark shape of the interface in Fig. \ref{fig:experiment_snapshots}. This is the consequence of the contact between the water and the glass walls. See Supplementary Information for a discussion on the role of this 3D effect.

\subsection{Liquid portion}

Pressure everywhere in the liquid is given by hydrostatic law:
\begin{equation}
p(Y) = p_a - \rho g (Y+H)
\label{eq:pressure}
\end{equation}
where $p_a$ is the pressure in the air, $\rho$ the density of the liquid and $g$ gravity acceleration (see Fig. \ref{fig:notations} for notations). Whenever $H \neq 0$, the liquid portion such that $Y>-H$ is in depression with respect to the atmosphere. 
Pressure $p(Y)$ jumps to $p_a$ across the curved liquid interface, due to Laplace law $\gamma \kappa =  \Delta p$. We introduce the angle $\psi$ between the tangent to the meniscus and the $X$-direction ($\psi$ is positive if trigonometric), which is related to the curvature by $\kappa = -\frac{\mathrm{d} \psi}{\mathrm{d} S} = -\psi '(S)$, $S$ being the arc-length along the meniscus. Laplace law can be written:
\begin{equation}
\gamma \, \psi ' (S) =  \rho g \, (Y_m(S) + H)
\end{equation}
where $Y_m$ is the vertical position of the liquid interface. Using the differential relation $Y_m'(S) = \sin \psi (S) $, we can finally write:
\begin{equation}
\psi '' (S) =  \frac{1}{L^2_\mathrm{gc}} \sin \psi (S).
\label{eq:meniscus}
\end{equation}
In the case of a meniscus vanishing far from the wall, $\psi(S~\to~\infty)~=~0$, equation (\ref{eq:meniscus}) has an analytical solution \citep{landau_elasticity}, which can be written:
\begin{equation} 
\psi(S) = 4 \arctan \left( \tan \frac{\psi_0}{4} \exp{ \frac{-S}{L_\mathrm{gc}} } \right) \,.
\label{eq:psi}
\end{equation}
Here $\psi_0$ is a constant, whose value is in general linked to the contact angle at the wall. In our system however, the contact angle changes with $H$, as already displayed in the previous section. We will show later the way in which $\psi_0$ can be found.


\subsection{Elastic portion}
\begin{figure}
    \centering
    \includegraphics[width=8.8cm]{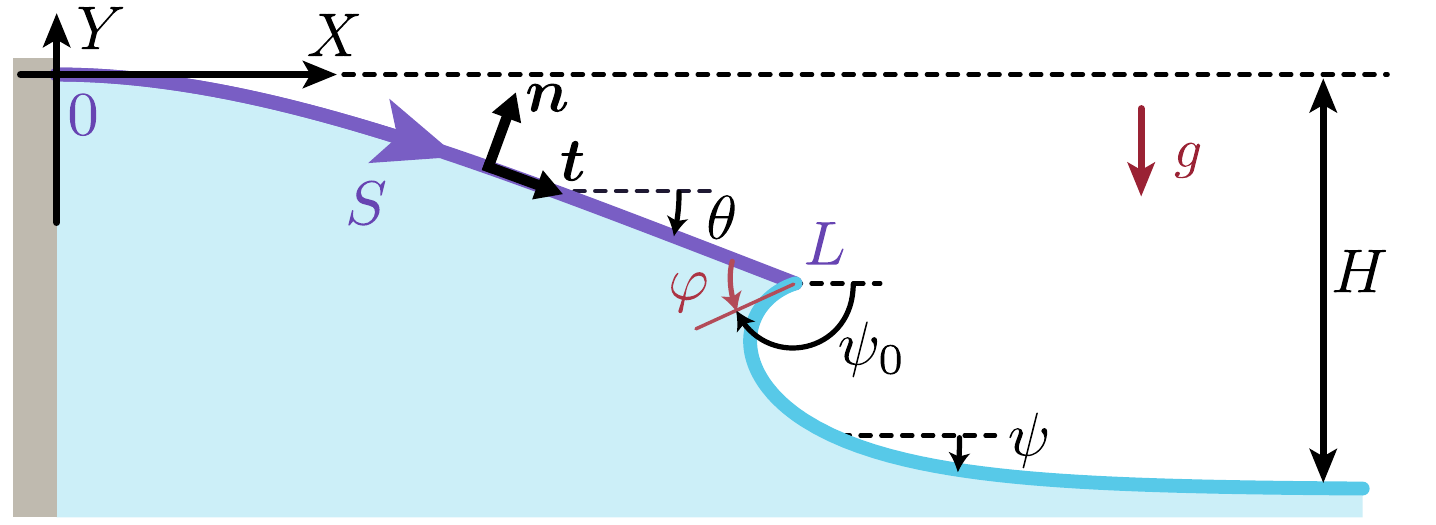}
   	 \caption{\textbf{Notations used in the model}. This sketch is a 2D view of Fig.~\ref{fig:experiment}, introducing the notations used in the model. The local angle between the $X$-axis and the strip (violet) is denoted $\theta$, the one between the $X$-axis and the free surface (blue) is $\psi$, and the contact angle between the liquid and the strip is $\varphi$.}
	 \label{fig:notations}
\end{figure}

The deformation of an inextensible beam can be expressed as a function of the angle $ \theta$ between the tangent to the beam and the $X$-direction ($\theta$ is positive if trigonometric). 

As the beam lies at the interface, it is subjected to air pressure on the upper face and water pressure on the lower face. 
Thus, the effective distributed force per unit width is $\mathbf{f} = -\rho g \, (Y+H) \, \mathbf{n}$, where $S$ is the arc-length along the beam, spanning from 0 to $L$, and $\mathbf{n} = - \sin \theta \mathbf{e}_x + \cos \theta \mathbf{e}_y$ is the unitary vector normal to the upper face. 
Note that we neglect the role of the own weight of the polymer strip, as it floats on the liquid interface.

Equilibrium of internal force $\mathbf{F}(S) = F_x(S) \mathbf{e}_x  + F_y(S) \mathbf{e}_y $ and internal moment $\mathbf{M}(S) = M(S) \, \mathbf{e}_z$  is given by Kirchhoff equations,
whose projections on the three axes are:
\begin{subequations}
\begin{align}
F'_x(S) &= - \rho g \, ( \, Y(S)+H) \sin \theta (S) \label{eq:kirch1} \\[.5em]
F'_y(S) &= \rho g \, (\, Y(S)+H) \cos \theta (S)  \label{eq:kirch2} \\[.4em]
B \theta''(S) &= F_x(S) \sin \theta (S) - F_y(S) \cos \theta (S)    \label{eq:kirch3}
\end{align}
\label{eq:kirchhoff}
\end{subequations}
These equations are written per unit width. We introduced the bending constitutive relation $B \theta '(S) = M(S)$, with $B = EI/W$ the bending stiffness per unit width ($E$ is the Young modulus and $I$ the second moment of the area of the cross section). 
The position of an infinitesimal element $\mathrm{d}s$ is related to the angle $\theta$ by the relations:
\begin{subequations}
\begin{align}
X'(S) &= \cos \theta (S) \label{eq:xprime_dim} \\[.5em]
Y'(S) &= \sin \theta (S) \label{eq:yprime_dim} \; .
\end{align}
\end{subequations}
Finally, boundary conditions are:
\begin{equation}
\begin{array}{rcr}
X(0) = 0 \; &;& \;  Y(0) = 0  \\[0.5em]
\theta (0) = 0 \; &;& \; \theta'(L) = 0
\end{array}
\label{eq:bc1}
\end{equation}
meaning that the strip is clamped horizontally at the origin and curvature vanishes at the end. Two more boundary conditions are related to the force applied by the liquid-air interface at the tip of the strip: here surface tension pulls the beam with a force of magnitude $\gamma$ and an angle $\psi_0$ with respect to the horizontal (see Fig. \ref{fig:notations}). Hence:
\begin{equation}
F_x(L) = \gamma \, \cos {\psi_0} \quad ; \quad F_y(L) = \gamma \, \sin {\psi_0} \, .
\label{eq:bc2psi}
\end{equation}
%
It is convenient to write these last boundary conditions as a function of the contact angle $\varphi$ between the solid and the liquid rather than  $\psi_0$. Looking at figure \ref{fig:notations} it appears that $\psi_0 = \theta(L) + \varphi - \pi$, therefore:
\begin{subequations}  
\begin{align}
F_x(L) &= - \gamma \, \cos \left( \theta(L) + \varphi \right) \\[.5em]
F_y(L) &= - \gamma \, \sin \left( \theta(L) + \varphi \right) \, .
\label{eq:bc2}
\end{align}
\end{subequations}
Kirchhoff equations (\ref{eq:kirchhoff}) can be written in a more compact form (see Supplementary Information for details):
\begin{equation}
B \theta'''(S) + \frac{B}{2} \theta'(S)^3   +  \gamma \theta'(S) \cos \varphi  + \rho g (Y(S) + H) = 0
\label{eq:main}
\end{equation}
 
A characteristic length scale appears naturally in this problem when a balance between elastic and hydrostatic forces applies, the elasto-hydrostatic length $L_\mathrm{eh} = (B/ \rho g)^{1/4}$. This length scale was already introduced by \citet{hertz1884} and \citet{foppl} in their founding works on elastic plates. The elasto-hydrostatic length is also the relevant length scale when considering the buckling of an elastic beam resting on a compliant substrate \citep{timoshenko}, in which the rigidity of the foundation is replaced by $\rho g$. 
%
We use $L_ \mathrm{eh}$ to make lengths dimensionless: 
\begin{equation}
\left\{s, x, y, h, \ell, \ell_\text{gc}\right\} = \frac{1}{L_\mathrm{eh}} \left\{S, X, Y, H, L, L_\mathrm{gc}\right\} 
\end{equation}

%
%
Dividing equation (\ref{eq:main}) by $\rho g L_{eh}$ leads to the system of equilibrium equations (dependence on $s$ in now suppressed for brevity):
\begin{subequations}
\begin{align}
\theta''' + \frac{1}{2} \theta'^3 + &\ell_\mathrm{gc} ^2 \cos \varphi \, \theta' + (y + h) = 0  \label{eq:theta_adim}\\[.2em]
x' &= \cos \theta  \label{eq:xprime_adim} \\[.5em]
y' &= \sin \theta  \label{eq:yprime_adim}
\end{align}
\label{eq:main_adim}
\end{subequations}
with boundary conditions:
\begin{equation}
\begin{array}{rcl}
x(0) = 0 & ;  & y(0) = 0 \quad;  \quad\theta(0) = 0  \\[.5em]
\theta'(\ell) = 0 &;& \theta''(\ell) = \ell_\mathrm{gc}^2 \sin \varphi 
\end{array}
\label{eq:bc_adim}
\end{equation} 
%



\subsection{Matching condition}

Because of contact line pinning at the edge, the contact angle $\varphi$ between the elastic strip and the liquid surface does not result from classical Young's construction. Rather, $\varphi$ is set by the geometrical constraint expressing the anchoring of the meniscus at the strip edge: 
\begin{equation}
y_\text{strip}(\ell) = y_\text{meniscus}(\ell). 
\label{eq:matching_condition}
\end{equation}
This sole condition does not guarantee a constant value for $\varphi$ and indeed, the contact angle exhibits experimentally a strong dependence with the non-dimensional depth $h$.
%

\noindent After integrating equation (\ref{eq:meniscus}) once, the right hand side of this condition can be rewritten as
\begin{equation}
y_\text{meniscus}(\ell) =  - 2 \ell_\mathrm{gc} \sin \frac{\psi_0}{2} - h \, ,
\end{equation}
leading to the following convenient end point condition for $\varphi$:
\begin{equation}
y_\text{strip}(\ell) =  2 \ell_\mathrm{gc} \cos \left( \frac{\varphi + \theta(\ell)}{2}  \right)  - h
\label{eq:findphi}
\end{equation}



\section{Elasto-capillary meniscus equilibrium shapes and their limit of existence}
\label{sec:max_height}

\begin{figure}
    \centering
    \setlength{\unitlength}{1cm}
    \begin{picture}(8.7,2.5)
    \put(0,0){\includegraphics[width=4.25cm]{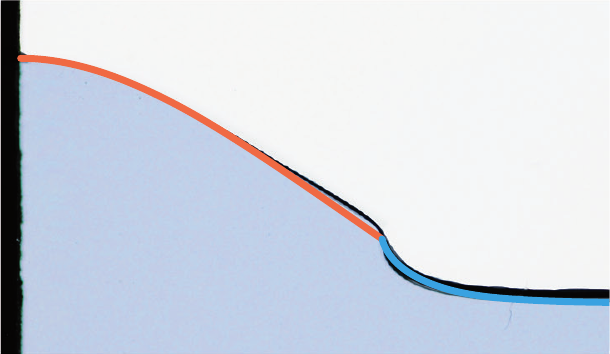}}
    \put(4.45,0){\includegraphics[width=4.25cm]{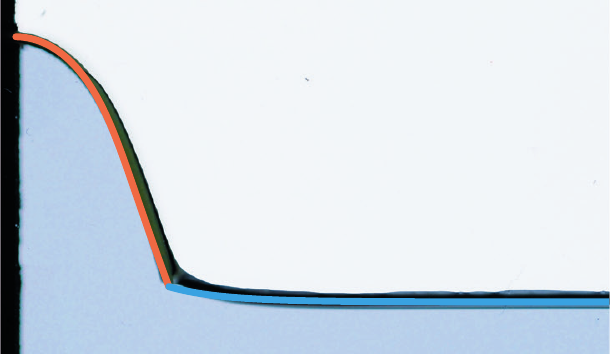}}
    \put(3.7,2.15){\textsf{(a)}}
    \put(8.1,2.15){\textsf{(b)}}
    \end{picture}
   	 \caption{\textbf{Comparison between experimental and theoretical profiles}. (a) The theoretical equilibrium shape (obtained with $\ell = 1.83$, $\ell_\mathrm{gc} = 0.24$ and $h = 1.09$) is superimposed to its experimental counterpart (corresponding to the last frame of Fig. \ref{fig:experiment_snapshots} panel a). (b) Same comparison with the second frame of Fig. \ref{fig:experiment_snapshots} panel b (here $\ell = 2.44$, $\ell_\mathrm{gc} = 0.48$ and $h = 2.06$).}
	 \label{fig:comparison_model_experiment}
\end{figure}

\subsection{Predicted and observed equilibrium shapes}

Typical equilibrium configurations observed in experiments for PET and PVS strips are reported Fig.~\ref{fig:comparison_model_experiment}. These two examples correspond to different values of the three dimensionless parameters ($h$, $\ell$ and $\ell_\mathrm{gc}$) governing the overall shape of the structure. 
In order to test the assumptions made in the model, we confront these observations with theoretical predictions. Specifically, we integrate numerically the equations (\ref{eq:main_adim}-\ref{eq:bc_adim}) associated with the matching condition (\ref{eq:findphi}). The shape predicted with the model for the experimental governing parameters is then superimposed to the observed profiles, and a good agreement can be noted both for PVS and PET strips.

Is the observed deflection of the strip gravity-driven or capillarity-driven? To answer this question we introduce the analogue of the elasto-hydrostatic length for capillarity, \emph{i.e.} the elasto-capillary length \citep{bico2004} $L_\mathrm{ec} = (B/ \gamma)^{1/2}$ which is the typical length needed for capillary bending effects to overcome the flexural rigidity of the strip. As $L_\mathrm{ec} L_\mathrm{gc} = L_\mathrm{eh}^2$ we see  that the nondimensional parameter $\ell_\mathrm{gc}$ can be written as the ratio $L_\mathrm{eh}/L_\mathrm{ec}$. Hence $\ell_\mathrm{gc}$ is a measure of the relative importance of hydrostatic forces over capillarity in the bending process. In our experiments, $\ell_\mathrm{gc}$ is systematically lower than $0.5$ but still of order unity, meaning that hydrostatic pressure is hardly the dominant effect in the deformation of the strip. Therefore we chose not to neglect $\ell_\mathrm{gc}$ in the following discussion. 

\subsection{The path towards extraction}

\begin{figure}
    \centering
    \includegraphics[width=8.7cm]{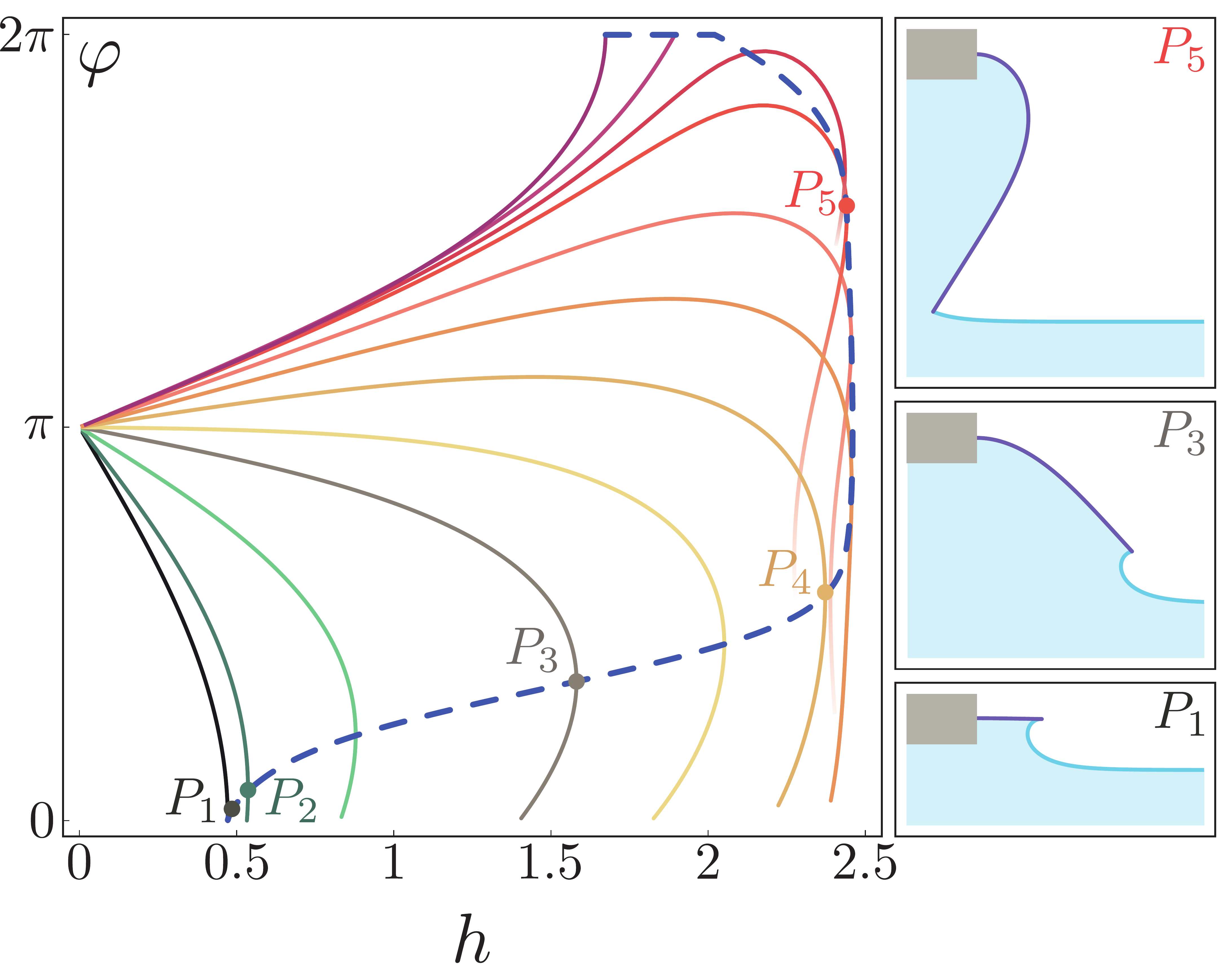}
    \caption{\textbf{Limit of existence for the elasto-capillary meniscus}. The paths followed by equilibrium states are shown in the $(h,\varphi)$ plane, in which $\varphi$ is restrained to physical values between 0 and $2 \pi$. The curves correspond to increasing values of the nondimensional strip length $\ell$, starting with a near zero value for the black curve (from bottom to top $\ell =$ 0.01, 1, 1.5, 1.8, 2, 2.2, 2.4, 2.6, 2.8, 2.9, 3, 3.2 respectively). It appears that for each curve a maximal value for $h$ exists, and beyond this limit point no more solution can be found. The dashed blue line reports the path of these limit points as $\ell$ is varied. Three equilibrium shapes, corresponding to the limit points $P_1$, $P_3$ and $P_5$, are reported on the right. Note that throughout this diagram, pinning of the contact line on the strip tip is assumed.}
    \label{fig:fold_pure_pinning}
\end{figure}

We now investigate the nature of the extraction process \emph{per se}, with a particular focus on the dependence of the critical height with the governing parameters. To this end, we make use of a continuation algorithm to follow the equilibrium configuration as the strip is withdrawn from the bath. These equilibria branches are reported Fig.~\ref{fig:fold_pure_pinning}. Note that for comparison purposes the value of the nondimensional capillary length has been set throughout this study to $\ell_\mathrm{gc}=0.24$, corresponding to the experiment involving a strip made of PET.

\noindent We start by considering the rigid limit case $\ell = 0.01$ where the
elasto-hydrostatic length $L_\text{eh}$ is 100 times larger than the length of
the strip $L$ (downmost black curve in Fig.\ref{fig:fold_pure_pinning}). When
$h = 0$, the liquid surface is perfectly horizontal ($\psi = 0$) and the
apparent contact angle is $\varphi = \pi$. As $h$ is increased, the contact
angle drops down to zero for $h = 2 \ell_\mathrm{gc} = 0.48$. In this rigid
case, $h$ simply measures the true height of the liquid meniscus. This height
cannot exceed the critical height $2 \ell_\mathrm{gc}$ corresponding to a
purely wetting state ($\varphi = 0$, see point $P_1$) -- otherwise the
matching condition~(\ref{eq:matching_condition}) would not be fulfilled. This
is precisely the critical height obtained by Laplace \citep{Laplace1805} that allowed
him to predict the critical extraction force, measured experimentally by Gay-Lussac.

Now adding some elasticity in the problem, we consider analogous equilibrium
branches for flexible strips ranging from $\ell=0.01$ up to $\ell=3.2$. At
first glance, it can be noticed than whenever $h=0$ the deflection of the strip is nil and so the contact angle
$\varphi$ is always equal to $\pi$. Starting from this flat
configuration, we follow each branch of equilibrium as $h$ varies. It is
noteworthy that the addition of elasticity allows to reach higher maximal
heights in the system. This is clearly illustrated by the elasto-capillary
menisci shapes at points $P_3$ and $P_5$ in Fig.~\ref{fig:fold_pure_pinning}.

Most of the equilibrium branches in Fig.~\ref{fig:fold_pure_pinning} exhibit a
limit (fold) point. In other words, two equilibria coexist for a given value
of $h$; presumably one stable and the other unstable. At the critical value,
these two equilibria coalesce and then disappear: \emph{there is no more
equilibrium solution past the limit point}. This behaviour is commonplace in
dynamical systems (saddle-node or fold bifurcation), and was first noticed in the context of interfaces by
Plateau \citep{Michael1981}. More precisely, when a catenoid made of soap film
is formed between two rigid rings, there is also a limit point where two
catenoids (one unstable and the other stable) merge and then cease to exist.
Experimentally, beyond the limit point, the catenoid bursts into droplets, and
the elasto-capillary meniscus breaks down (\emph{i.e.} the lamella is
extracted) as illustrated Fig.~\ref{fig:collapse_classique}. Note that for the
two uppermost curves ($\ell = 3$ and $\ell = 3.2$) the maximum value for $h$
no more corresponds to a limit point, but rather to a constrained maximum
occurring when $\varphi = 2 \pi$, the largest possible value for the contact
angle. 

The locus of these maxima, be them limit fold points or constrained maxima, is
revealed Fig.~\ref{fig:fold_pure_pinning} with a dashed blue line. The path
followed by these maxima appears non-trivial, suggesting optimum values for
the elasticity to achieve largest heights and exhibiting a tumble as $\varphi$
approches $2 \pi$. As these points mark the frontier where equilibria
disappear, it is therefore quite natural to expect from this line to represent
the \emph{extraction limit} for the strip.

\subsection{Pinning and slipping of the contact line}

\begin{figure}
    \centering
    \includegraphics[width=8.7cm]{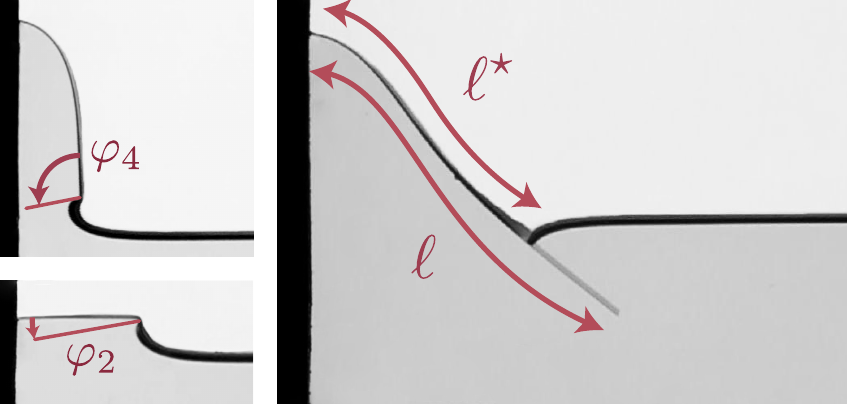}
    \caption{\textbf{Slipping of the contact line}. Top left: experimental shape observed just before air invasion sets in. The strip has a nondimensional length $\ell = 2.2$, whose limit point in Fig.~\ref{fig:fold_pure_pinning} is the point $P_4$. The superimposed angle is the contact angle reached at the limit point $P_4$, and it is in good agreement with the one observed in the experiment. Bottom left: snapshot corresponding to a strip with $\ell = 1$. Here, air invasion is triggered for a contact angle $\varphi$ clearly above $\varphi_2$, the contact angle associated with point $P_2$. The instability involves a slipping of the contact line as soon as the angle reaches some critical value $\varphi_c$. Right: Equilibrium state for $\ell = 3.68$. The strip literally plunges in the bath, with an effective `interfacial length' $\ell^\star < \ell$.}
    \label{fig:slipping}
\end{figure}

A simple way to verify the exactitude of the extraction prediction is to
compare the theoretical shape of the elasto-capillary meniscus to its
corresponding experimental counterpart.

Fig.~\ref{fig:slipping} (left) we show two experimental pictures taken at the
moment of the extraction. In the first case ($\ell=2.2$ -- top left image),
the experimental contact angle at extraction accurately matches the angle
$\varphi_4$ expected from the theory ($\varphi_4$~corresponds to the angle at
point $P_4$ in Fig.~\ref{fig:fold_pure_pinning}). This means that the folding
point $P_4$ accurately captures the threshold for extraction, comforting our
previous analysis. However, the second case shown Fig.~\ref{fig:slipping}
($\ell =1.0$ -- bottom left) exhibits a poor agreement with the theoretical
prediction. There, the experimental critical contact angle is much larger than
the value $\varphi_2$ expected from the location of folding point $P_2$ in
Fig.~\ref{fig:fold_pure_pinning}.

The reason for this discrepancy lies in the contact line pinning hypothesis,
that proves to be faulty for some values of the contact angle. Indeed,
experimentally we clearly observe an advance of the contact line as soon as
$\varphi$ approaches a critical value $\varphi_c \simeq 80^\circ$. The
slippage of the contact line, associated with air invasion underneath the
elastic strip, is an unstoppable process; it proceeds until the lamella is
fully extracted.

Analogously, when the free surface overturns the strip tip a similar advance
of the contact line can be observed. Actually it is triggered as soon as the
air wedge angle is less than $\varphi_\text{c}$, that is, when $\varphi > 2\pi
- \varphi_\text{c}$. But this time, the slippage is not necessarily a runaway
process leading to extraction, as exemplified~Fig.~\ref{fig:slipping} (right)
with a \emph{stable} configuration.  There, water invasion sets in as soon as
the critical angle $2\pi - \varphi_\text{c}$ is reached. But the system now
settles to a new `plunged state' that reduces the effective `interfacial
length' of the strip $\ell^\star (< \ell)$. In this type of configuration,
part of the strip remains stably underwater.

These experimental observations tend to depict a somewhat more complex picture
of the extraction process than previously expected. Specifically, the locus of
the folding points in Fig. \ref{fig:fold_pure_pinning} does not necessarily
describe the extraction threshold. Extraction cannot be grasped without paying
attention to the slipping motions of the contact line on the upper or lower
side of the strip.
    
\begin{figure}
    \centering
    \includegraphics[width=8.8cm]{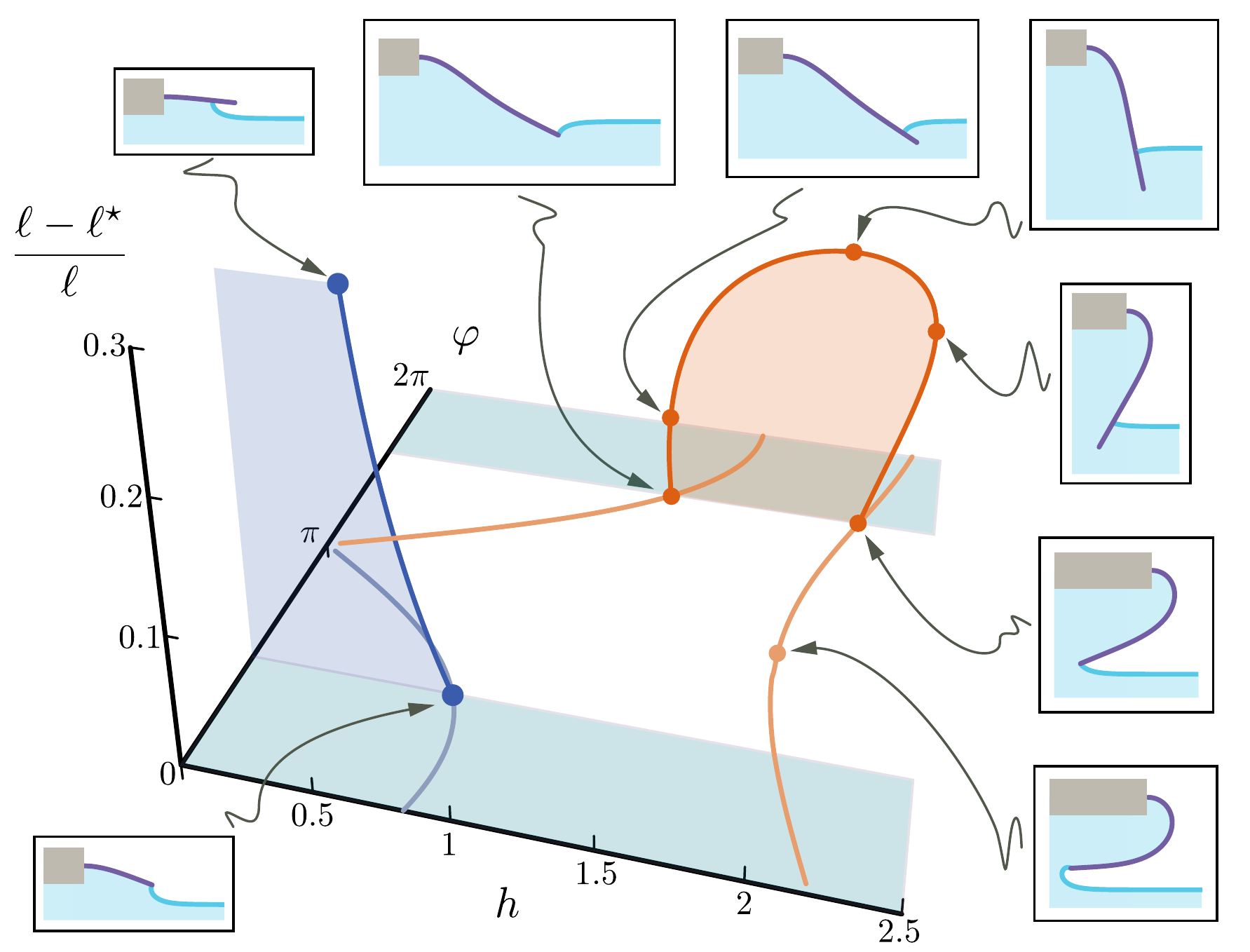}
    \caption{\textbf{Equilibrium branches including pin-slip effects}. Loci of equilibria in the ($h$,$\varphi$,$(\ell-\ell^\star)/\ell$) space are shown for $\ell=1.5$ (blue line) and $\ell = 3.4$ (orange line). As long as the contact angle $\varphi$ lies in the range $(\varphi_c,2\pi-\varphi_c)$ (i.e. between the blue stripes), the contact line is pinned to the strip tip and equilibrium branches lie in the $(h,\varphi)$ plane. When the contact angle reaches a critical value a bifurcation occurs. There, another branch corresponding to slipping contact line takes off out-of-plane, as the interfacial length $\ell^\star$ no more corresponds to the actual length of the strip $\ell$. Reconnection to pinned states appears to be possible.}
    \label{fig:plot3D}
\end{figure}

Contact line slippage can however be included in the model\citep{Rivetti2012}. Indeed, assuming
that the contact angle is constant during slipping\citep{gennes1985} (and
equal to $\varphi_\text{c}$ or $2 \pi - \varphi_\text{c}$), we can still
search for equilibrium solutions provided that the interfacial length
$\ell^\star$ is turned into a free parameter. Note that the total number of
unknowns in this setting is unchanged as $\varphi$ is known. The end
condition~(\ref{eq:findphi}) can now be used as a condition for
$\ell^\star$, and the system of leading equations
(\ref{eq:main_adim}-\ref{eq:bc_adim}) be integrated from $0$ to $\ell^\star$.

Figure~\ref{fig:plot3D} displays the equilibrium branches for two values of
the nondimensional length, $\ell =1.5 $ and $\ell = 3.4$, when both pinning
and slipping of the contact line are retained. These branches are plotted in
the three-dimensional space $(h,\varphi,(\ell-\ell^*)/\ell)$. Here
$(\ell-\ell^*)/\ell$ represents the relative immersed/emerged portion of the
strip (this portion can either be in air or water). Note that this quantity
cancels out when the contact line is pinned to the strip edge.

For $\ell=1.5$ (blue line), the path followed by equilibrium solutions is
identical to the one already shown Fig.~\ref{fig:fold_pure_pinning} for pure
pinning. However, as the contact angle $\varphi$ reaches the critical value
$\varphi_\text{c}$ an out-of-plane bifurcation occurs. The new branch arises
from contact line slipping along the underside of the strip, as shown in the
inset. This bifurcation is subcritical and the equilibrium branch
monotonically tends to $(\ell - \ell^\star)/\ell = 1$, meaning full extraction
of the strip (for clarity reasons, the branch is only displayed for values
such that $(\ell - \ell^\star)/\ell < 0.3$). A careful investigation of the
nature of the bifurcations occurring at $\varphi = \varphi_\text{c}$ suggests
that those are all subcritical, \emph{i.e.} lead to a sudden extraction of
the strip. 

For $\ell = 3.4$ (orange line) the contact angle now increases with $h$. The
free surface overturns the strip edge, and a slipping bifurcation occurs when
$\varphi = 2\pi - \varphi_c$. The nature of the bifurcation is now different,
as equilibrium `plunged states' exist for values of $h$ larger than that at
the bifurcation. Typical shapes of plunged states are shown in the insets of
Fig.~\ref{fig:plot3D} and illustrate the slipping of the contact line above
the strip. The loop shape of the out-of-plane branch implies that the
interfacial length $\ell^\star$ first decreases with $h$ until it reaches a
minimum value, and increases up to $\ell$ again as shown in the insets. In
contrast to air invasion the immersion of the strip is thus a reversible
phenomenon. Surprisingly enough, the slipped states branch then exhibits a
reconnection with a new in-plane (pinned states) branch. It is remarkable to
point out that the two in-plane orange branches are by no way connected, even
when $\varphi$ is allowed to adopt values far beyond its physical limits (data
not shown). This is the ambivalent `pin-slip' nature of the end condition that
allows for such a plentiful bifurcation diagram allowing to track equilibria
along multiple branches, at variance with the catenoid case where the end
condition is unique.

\begin{figure}     
	\centering
	\includegraphics[width=8.8cm]{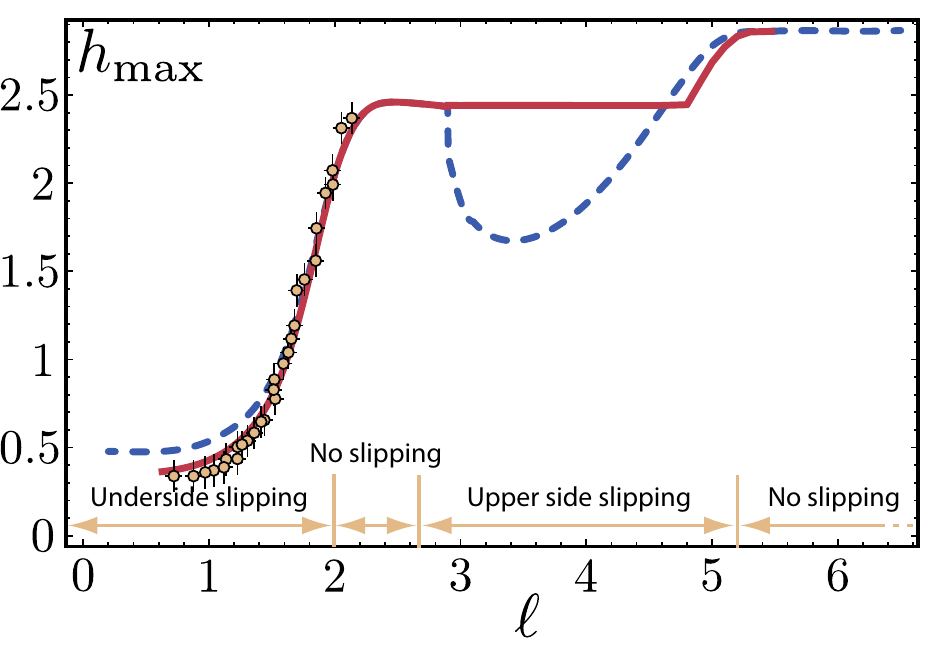}     
	\caption{\textbf{Critical meniscus height versus strip length}. 
		Comparison between the critical height
		$h_\text{max}$ reached in experiments as a function of the nondimensional
		strip length $\ell$ (orange dots) and the theoretical prediction based on
		the `pin-slip' end condition (magenta curve). For reference, the results using a 
		pure pinning end condition are
		shown with the dashed blue curve. The different r\'egimes in terms of slipping
		of the contact line are also displayed.}     
	\label{fig:hmaxvsl}
\end{figure}

We now confront the experimental critical height $h_\mathrm{max}$ reached by
the elasto-capillary meniscus as a function of $\ell$ with the model,
considering both pure pinning and `pin-slip' conditions
(Fig.~\ref{fig:hmaxvsl}) (in the pure pinning case the dashed curve reporting
the critical height is basically a translation of the limit points locus shown
Fig.~\ref{fig:fold_pure_pinning} in a function of $\ell$). Clearly the
agreement between the experimental observations and the `pin-slip' prediction
is excellent. The differences between pure pinning and pin-slip are even more
pronounced for larger values of $\ell$. Indeed, for pure pinning a strange tumbling behaviour is
particularly noticeable in the region $3<\ell<5$, suggesting the odd idea that
lengthening the strip shortens the critical height. This wrong prediction is
actually linked to the non-monotonic shape of the limit points locus. The full
model incorporating mixed `pin-slip' end conditions allows for a correction of
this prediction and smoothes out the evolution of $h_\text{max}$ with $\ell$.
This highlights the importance of considering out-of-plane bifurcations and
reconnections inherent to contact line slipping along the upper side of the
strip in the extraction process. Note also that Fig.~\ref{fig:hmaxvsl}
exhibits the whole r\'egime phenomenology appearing in this problem, along
with their non-trivial apparition sequence. Finally, looking at
Fig.~\ref{fig:hmaxvsl} suggests an asymptotic behavior for $\ell \gg 1$, a
possibility we investigate next.

\subsection{Universal behavior of long strips}

We now present a theoretical focus on the case $\ell \gg 1$ where the strip is
very long compared to the elasto-hydrostatic length. Rescaling
(\ref{eq:yprime_adim}) and (\ref{eq:theta_adim}) with $\tilde{s} = s/\ell$
and $\tilde{y} = y/h$ leads to:
\begin{subequations}
\begin{equation} 
\frac{h}{\ell} \tilde{y} '(\tilde{s}) = \sin \theta ( \tilde{s} ) 
\label{eq:yprime_ext}
\end{equation} 
\begin{equation}   
1 +\tilde{y}(\tilde{s}) +\frac{1}{\ell ^3 h} ( \theta ''' (\tilde{s})+
\frac{1}{2} {\theta'(\tilde{s})} ^3 ) +  \frac{\ell _\mathrm{gc}^2}{\ell h }
\theta ' (\tilde{s}) \cos \varphi =0.
\label{eq:main_ext}
\end{equation}  
\end{subequations}
Taking $1/\ell$ as a small parameter, we immediately obtain at leading order the external
solution $\theta=0$ and $\tilde{y} = -1$. This external solution represents a
strip lying flat over the liquid surface, with a contact angle $\varphi=\pi$.
A boundary layer solution exists and reconciles the external flat solution
with the boundary condition $\tilde{y} (0) = 0$. We introduce classically the
boundary layer variable $\tilde S$ as follows: $\tilde{s} = \epsilon
\tilde{S}$, with $\tilde{S} = O(1)$ within the boundary layer. In terms of
this variable, equation~(\ref{eq:yprime_ext}) reads:
\begin{equation} \frac{h}{\ell}
\frac{1}{\epsilon} \tilde{y}' (\tilde{S}) = \sin \theta   (\tilde{S}),
\end{equation}
and dominant balance requirements immediately give the thickness of the boundary layer
as $\epsilon\sim~h/\ell$.  

Injecting this results into equation (\ref{eq:main_ext}) yields:
\begin{equation} 
\frac{1}{h^4} \left( \theta ''' (\tilde{S}) +
\frac{1}{2}{\theta'}^3 (\tilde{S}) \right)  +  (\tilde {y} (\tilde{S}) + 1) -
\frac{\ell_ {gc}^2}{h^2} \theta ' (\tilde{S})  = 0 
\label{eq:bound_layer}
\end{equation}

In this equation the hydrostatic pressure term is of order one, and has to be
balanced by -- at least -- one other term. Let's first consider the case
$\ell_\mathrm{gc} \ll h$, where hydrostatic pressure is solely balanced by
elasticity. The balance can only happen if $h \sim 1$, consistently with the
significance of the elasto-hydrostatic length $L_\text{eh}$. This constraint
certainly gives clues on the asymptotic bounded behaviour of $h_\mathrm{max}$
when $\ell$ achieves large values, as shown Fig.~\ref{fig:hmaxvsl}. In this limit (amounting to neglect capillary forces) the
shape of the strip is governed by the following equation -- expressed with the original variables: 
\begin{equation} 
\theta ''' + \frac{1}{2}{\theta'}^3   +  y + h = 0.
\label{eq:bound_layer_elast} 
\end{equation} %

A second option in the dominant-balance analysis is to retain only hydrostatic
pressure effects and surface tension ones. This corresponds typically to $h
\sim \ell_\mathrm{gc}$ and $\ell_\mathrm{gc} \gg 1$. The shape adopted by the
strip in this r\'egime is described by solutions of:
\begin{equation}
y + h - \ell_ {gc}^2 \theta ' = 0.
\label{eq:bound_layer_cap}
\end{equation}

When derived this equation can be recast into the well known meniscus
equation $\theta '' = 1/\ell_ {gc}^2 \sin \theta$. The shape of the elastic
strip is thus that of a pure capillary meniscus. Actually this is nearly the
case, as solutions of this equation cannot meet the clamped boundary
conditions at the wall, as equation~(\ref{eq:bound_layer_cap}) has lost two
derivation orders with respect to equation~(\ref{eq:bound_layer}). A third
layer, in which elasticity is restored, can however be introduced to fulfill
the correct boundary conditions of the problem.

Eventually, an appealing feature of the full equation~(\ref{eq:bound_layer})
should be pointed out:  its structure bears a striking similarity with the
equilibrium equation of a buckled elastic beam sitting on a liquid foundation,
for which an analytical solution exists \citep{diamant2011}. We have recently
demonstrated that the exact solution of equation (\ref{eq:bound_layer}) can be
obtained as well and is generally not symmetric \citep{rivetti_cras}.



\section{Extraction force}
\label{sec:force}

%


We are now in a position to address the main question raised in this study: How does elasticity affect the extraction process?
We start by introducing the (dimensionless) pulling force $f_\text{p}$, that is the vertical component of the force exerted by the clamp: 
\begin{equation}
f_\text{p} = -f_y (0) = \int_0^\ell (y(s)+h) \cos \theta (s) \, \mathrm{d} s \,- \ell^2_\mathrm{gc} \sin \psi_0
\end{equation}
Although the pulling force mainly balances the weight of the volume of fluid displaced by the strip, it does rely intimately on elasticity and capillarity as these effects shape the liquid volume.
As  the imposed height $h$ is increased, the lifted liquid volume varies, and so does the pulling force $f_\text{p}$. The \textit{extraction force} $f_\mathrm{ext}$ is therefore defined as the largest value of $f_\text{p}$ with respect to $h$: $f_\mathrm{ext}= \mathrm{max}_h \,f_\text{p}(h)$. Note that in general this maximal value is not achieved when $h = h_\mathrm{max}$.


\begin{figure}
    \centering
    \includegraphics[width=8.8cm]{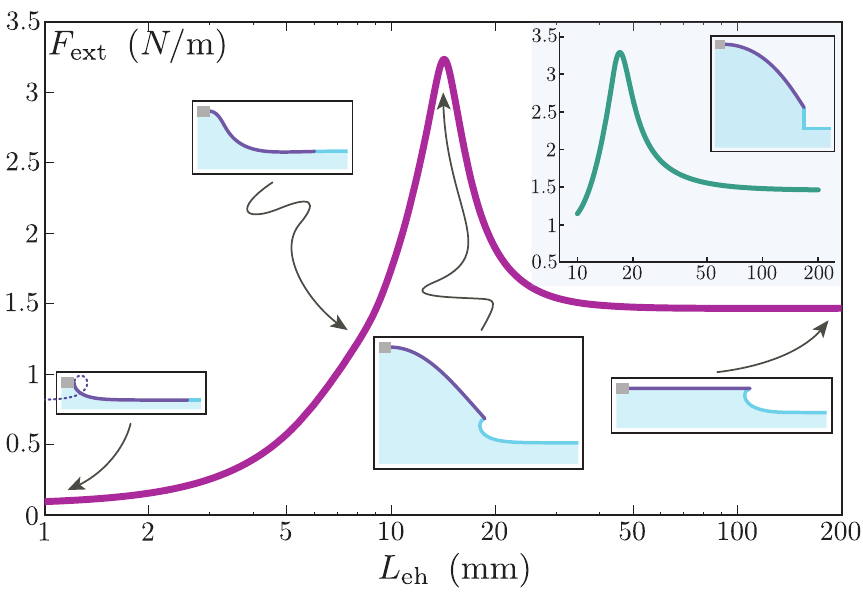}
    \caption{\textbf{Extraction force versus elasto-hydrostatic length.} Relation between the dimensional extraction force per unit width $F_\text{ext}$ (in $N/$m) and the rigidity of a fixed-length strip, here measured in terms of the elasto-hydrostatic length $L_\text{eh}$ (mm). The length of the strip taken here is 27 mm and the gravito-capillary length $L_\text{gc}$ has been set to 2.7 mm. The figure clearly displays an adhesion peak  for a specific value of $L_\text{eh}$ around 0.5 $L$. Representative  snapshots of the elasto-capillary meniscus at the moment of extraction are disposed along the curve. These snapshots correspond to different r\' egimes; From left to right: capillary limit, elasto-hydrostatic r\'egime, optimal adhesion, rigid limit. In the inset, a rudimentary model where the strip is approximated with a linearly varying curvature rod and the meniscus with a vertical liquid wall proves to capture the main features of the main graph, both qualitatively and quantitatively.}
    \label{fig:force_vs_leh}
\end{figure}
In order to investigate the exact role of elasticity in the extraction process, we represent Fig.~\ref{fig:force_vs_leh} the variation of the force needed to pull out a fixed-length strip with variable rigidity. More precisely the dimensional extraction force per unit width  $F_\mathrm{ext} = \rho g L^2_\mathrm{eh} f_\mathrm{ext}$ is represented as a function of $L_\mathrm{eh}$ for a given strip length $L$. For plotting purposes, we chose a strip  length of $L = 27$ mm and a fixed gravito-capillary length $L_\mathrm{gc} = 2.7$ mm corresponding to our experiments, but the features displayed Fig.~\ref{fig:force_vs_leh} appear to be generic. 

Different r\'egimes in the extraction process can readily be noticed on Fig.~\ref{fig:force_vs_leh} inspection. First, for very small values of $L_\text{eh}$ (corresponding to $L_\mathrm{eh} < L_\mathrm{gc}$, that is $\ell_\mathrm{gc} > 1$), the strip is so soft that it deforms almost with no resistance under the combined action of capillarity and gravity, as would have a pure liquid interface. This capillary limit is emphasized in the leftmost thumbnail of Fig.~\ref{fig:force_vs_leh} with a comparison between the elasto-capillary meniscus shape and the looping shape of a real liquid meniscus. This excellent agreement is consistent with the fact that in the limit  $\ell_\mathrm{gc} > 1$, the elasto-capillary meniscus shape is captured by equation~(\ref{eq:bound_layer_cap}).

When $L_\text{eh}$ reaches slightly larger values (but still such that $L_\text{eh} \ll L$, meaning $\ell \gg 1$), elasticity comes into play with the elasto-hydrostatic r\'egime. There, the strip is flat and aligned with the liquid surface outside a region of size $\mathcal O(H/L_\text{eh})$ where the deformation is concentrated, as already discussed. The shape of the elasto-capillary meniscus in the inner region is given by solving equation~(\ref{eq:bound_layer_elast}). Note that as long as $\ell$ is still larger than unity, the shape of the strip exhibits a self-similar behavior as it only depends on the elasto-hydrostatic length $L_\text{eh}$, and so does the lifted liquid volume: $F_\text{ext} \sim L_\text{eh}^2$ (\textit{i.e.} $f_\text{ext}$ constant). Stiffening the elastic strip therefore results in a quadratic increase of the extraction force.   

For higher values of $L_\text{eh}$ an adhesion peak clearly enters into the picture. This peak, centered at $L_\mathrm{eh} \simeq 0.5 L$ (\textit{i.e.} $\ell \simeq 2$), corresponds  to a cut-off of the self-similar r\'egime imposed by end-effects, as $L_\text{eh}$ becomes of the same order of magnitude as $L$. Here, the large deflection of the strip cause large amount of liquid to be drawn behind, hence a maximal extraction force.
For even larger values of $L_\mathrm{eh}$, the elastic strip progressively rigidifies as the extraction force falls down to its asymptotic Laplace/Gay-Lussac limit. It should be pointed out that the extraction work, relevant in the context of imposed displacement, displays a similar adhesion peak (data not shown).

This sequence of different r\'egimes can be captured with a rough model containing as only ingredients elasticity and gravity (inset of Fig.~\ref{fig:force_vs_leh}). In this model, the curvature of the strip is approximated with a linear function ramping from  $H/L_\mathrm{eh}^2$ at the wall down to 0 at the tip. The liquid meniscus touching the strip tip is in turn crudely approximated with a vertical liquid wall (see snapshot in the inset).
Finding the extraction force amounts to a maximization problem: For a given $L_\mathrm{eh}$, we look for the value of $H$ that maximizes the liquid volume risen by the strip, taking care that the height of the liquid wall cannot exceed $2L_\mathrm{gc}$. This simple model, disregarding both surface tension forces and nonlinear geometric displacements of the strip, recovers nicely the main features of the relation $F_\mathrm{ext}$ vs $L_\mathrm{eh}$: the adhesion peak, the different limits, the optimal shape. The quantitative agreement of this elementary model, free of adjustable parameters, certainly reveals the key physical ingredients necessary to understand the role of elasticity in capillary adhesion.

\section{Conclusion}

Elasticity is known to increase energetic costs in adhesion phenomena, and to induce hysteric cycles with different deformations back and forth. For example, the long polymer chains in elastomers can store all the more elastic energy that the chain is long. This energy is lost when the chains rupture and that cracks initiate \citep{Lake1967}. Analogously, soap bubbles gently put into contact can be deformed and remain connected even when the imposed separation exceeds the initial critical adhesion distance \citep{Besson2007}, thus implying bubbles conformations unexplored previously. 

In this study, we have focused on a paradigm setup coupling elasticity and capillary adhesion. The elastocapillary meniscus discussed here exhibits all the typical features of elastic adhesives: extraction force/work possibly increased by elasticity, hysteric cycles with deformations of the strip depending on whether it touches the interface or not, elastic energy stored released violently at extraction. This last observation, documented in Fig.~\ref{fig:collapse_classique} and in supplementary material, is also reminiscent of cavitation pockets appearing in the traction of adhesive films\citep{Poivet2003}.

A key point of our analysis is to highlight the existence of optimal strip stiffness corresponding to a maximal extraction force (conversely, this force can be as low as the material is soft). This could be exploited for example to improve the design of soft micro-manipulators or MEMS such as elastopipettes\citep{reis2010a}. Though the maximal volume displaced by the strip does not correspond to the actual amount of liquid grabbed from an interface, the extraction force and work do have an important role in this problem as they set the limit points of extraction. 

Finally, the agreement between theoretical and experimental shapes of an elasto-capillary meniscus, along with their dependence with the elasto-hydrostatic length $L_\text{eh}$,  suggests new metrology tools for a precise measure of the elastic properties of soft films.

\subsection*{Acknowledgments}
The authors are grateful to Sébastien Neukirch for constant encouragements and insights into dynamical systems, Cyprien Gay for suggesting many interesting references in adhesion phenomena, Pedro Reis for indicating historical references on elastic beams resting on soft foundations, Emmanuel de Langre, Jérôme Hoepffner and Andrés A. Le\'{o}n Baldelli for helpful discussions, Laurent Quartier for his help in the setup building. L’Agence Nationale de la Recherche through its Grant ‘‘DEFORMATION’’ ANR-09-JCJC-0022-01 and the Émergence(s) program of the Ville de Paris are acknowledged for their financial support.

\bibliographystyle{rsc} 
\bibliography{elastic_meniscus}

\providecommand*{\mcitethebibliography}{\thebibliography}
\csname @ifundefined\endcsname{endmcitethebibliography}
{\let\endmcitethebibliography\endthebibliography}{}
\begin{mcitethebibliography}{30}
\providecommand*{\natexlab}[1]{#1}
\providecommand*{\mciteSetBstSublistMode}[1]{}
\providecommand*{\mciteSetBstMaxWidthForm}[2]{}
\providecommand*{\mciteBstWouldAddEndPuncttrue}
  {\def\EndOfBibitem{\unskip.}}
\providecommand*{\mciteBstWouldAddEndPunctfalse}
  {\let\EndOfBibitem\relax}
\providecommand*{\mciteSetBstMidEndSepPunct}[3]{}
\providecommand*{\mciteSetBstSublistLabelBeginEnd}[3]{}
\providecommand*{\EndOfBibitem}{}
\mciteSetBstSublistMode{f}
\mciteSetBstMaxWidthForm{subitem}
{(\emph{\alph{mcitesubitemcount}})}
\mciteSetBstSublistLabelBeginEnd{\mcitemaxwidthsubitemform\space}
{\relax}{\relax}

\bibitem[de~{Laplace}(1805)]{Laplace1805}
P.~S. de~{Laplace}, \emph{Trait{\'e} de m{\'e}canique c{\'e}leste.}, Courcier,
  Paris, 1805, Suppl{\'e}ment au livre X\relax
\mciteBstWouldAddEndPuncttrue
\mciteSetBstMidEndSepPunct{\mcitedefaultmidpunct}
{\mcitedefaultendpunct}{\mcitedefaultseppunct}\relax
\EndOfBibitem
\bibitem[Nicolson(1949)]{Nicolson1949}
M.~M. Nicolson, \emph{Math. Proc. Cambridge Philos. Soc.}, 1949, \textbf{45},
  288--295\relax
\mciteBstWouldAddEndPuncttrue
\mciteSetBstMidEndSepPunct{\mcitedefaultmidpunct}
{\mcitedefaultendpunct}{\mcitedefaultseppunct}\relax
\EndOfBibitem
\bibitem[Srinivasan \emph{et~al.}(2001)Srinivasan, Liepmann, and
  Howe]{Srinivasan2001}
U.~Srinivasan, D.~Liepmann and R.~Howe, \emph{J. Microelectromech. Syst.},
  2001, \textbf{10}, 17--24\relax
\mciteBstWouldAddEndPuncttrue
\mciteSetBstMidEndSepPunct{\mcitedefaultmidpunct}
{\mcitedefaultendpunct}{\mcitedefaultseppunct}\relax
\EndOfBibitem
\bibitem[Botto \emph{et~al.}(2012)Botto, Lewandowski, Cavallaro, and
  Stebe]{botto2012}
L.~Botto, E.~P. Lewandowski, M.~Cavallaro and K.~J. Stebe, \emph{Soft Matter},
  2012, \textbf{8}, 9957--9971\relax
\mciteBstWouldAddEndPuncttrue
\mciteSetBstMidEndSepPunct{\mcitedefaultmidpunct}
{\mcitedefaultendpunct}{\mcitedefaultseppunct}\relax
\EndOfBibitem
\bibitem[Bico \emph{et~al.}(2004)Bico, Roman, Moulin, and Boudaoud]{bico2004}
J.~Bico, B.~Roman, L.~Moulin and A.~Boudaoud, \emph{Nature}, 2004,
  \textbf{432}, 690--690\relax
\mciteBstWouldAddEndPuncttrue
\mciteSetBstMidEndSepPunct{\mcitedefaultmidpunct}
{\mcitedefaultendpunct}{\mcitedefaultseppunct}\relax
\EndOfBibitem
\bibitem[Neukirch \emph{et~al.}(2007)Neukirch, Roman, de~Gaudemaris, and
  Bico]{neukirch2007}
S.~Neukirch, B.~Roman, B.~de~Gaudemaris and J.~Bico, \emph{Journal of the
  Mechanics and Physics of Solids}, 2007, \textbf{55}, 1212 -- 1235\relax
\mciteBstWouldAddEndPuncttrue
\mciteSetBstMidEndSepPunct{\mcitedefaultmidpunct}
{\mcitedefaultendpunct}{\mcitedefaultseppunct}\relax
\EndOfBibitem
\bibitem[Huang \emph{et~al.}(2007)Huang, Juszkiewicz, de~Jeu, Cerda, Emrick,
  Menon, and Russell]{huang2007}
J.~Huang, M.~Juszkiewicz, W.~H. de~Jeu, E.~Cerda, T.~Emrick, N.~Menon and T.~P.
  Russell, \emph{Science}, 2007, \textbf{317}, 650--653\relax
\mciteBstWouldAddEndPuncttrue
\mciteSetBstMidEndSepPunct{\mcitedefaultmidpunct}
{\mcitedefaultendpunct}{\mcitedefaultseppunct}\relax
\EndOfBibitem
\bibitem[Antkowiak \emph{et~al.}(2011)Antkowiak, Audoly, Josserand, Neukirch,
  and Rivetti]{Antkowiak2011}
A.~Antkowiak, B.~Audoly, C.~Josserand, S.~Neukirch and M.~Rivetti, \emph{Proc.
  Natl. Acad. Sci. U.S.A.}, 2011, \textbf{108}, 10400--10404\relax
\mciteBstWouldAddEndPuncttrue
\mciteSetBstMidEndSepPunct{\mcitedefaultmidpunct}
{\mcitedefaultendpunct}{\mcitedefaultseppunct}\relax
\EndOfBibitem
\bibitem[Blow and Yeomans(2010)]{blow2010}
M.~L. Blow and J.~M. Yeomans, \emph{Langmuir}, 2010, \textbf{26},
  16071--16083\relax
\mciteBstWouldAddEndPuncttrue
\mciteSetBstMidEndSepPunct{\mcitedefaultmidpunct}
{\mcitedefaultendpunct}{\mcitedefaultseppunct}\relax
\EndOfBibitem
\bibitem[Roman and Bico(2010)]{roman2010}
B.~Roman and J.~Bico, \emph{Journal of Physics: Condensed Matter}, 2010,
  \textbf{22}, 493101\relax
\mciteBstWouldAddEndPuncttrue
\mciteSetBstMidEndSepPunct{\mcitedefaultmidpunct}
{\mcitedefaultendpunct}{\mcitedefaultseppunct}\relax
\EndOfBibitem
\bibitem[Armstrong(2002)]{armstrong2002}
J.~E. Armstrong, \emph{American Journal of Botany}, 2002, \textbf{89},
  362--365\relax
\mciteBstWouldAddEndPuncttrue
\mciteSetBstMidEndSepPunct{\mcitedefaultmidpunct}
{\mcitedefaultendpunct}{\mcitedefaultseppunct}\relax
\EndOfBibitem
\bibitem[Reis \emph{et~al.}(2010)Reis, Hure, Jung, Bush, and Clanet]{reis2010a}
P.~M. Reis, J.~Hure, S.~Jung, J.~W.~M. Bush and C.~Clanet, \emph{Soft Matter},
  2010, \textbf{6}, 5705--5708\relax
\mciteBstWouldAddEndPuncttrue
\mciteSetBstMidEndSepPunct{\mcitedefaultmidpunct}
{\mcitedefaultendpunct}{\mcitedefaultseppunct}\relax
\EndOfBibitem
\bibitem[Holmes and Crosby(2010)]{Holmes2010}
D.~P. Holmes and A.~J. Crosby, \emph{Phys. Rev. Lett.}, 2010, \textbf{105},
  038303\relax
\mciteBstWouldAddEndPuncttrue
\mciteSetBstMidEndSepPunct{\mcitedefaultmidpunct}
{\mcitedefaultendpunct}{\mcitedefaultseppunct}\relax
\EndOfBibitem
\bibitem[Andreotti \emph{et~al.}(2011)Andreotti, Marchand, Das, and
  Snoeijer]{andreotti2011}
B.~Andreotti, A.~Marchand, S.~Das and J.~H. Snoeijer, \emph{Phys. Rev. E},
  2011, \textbf{84}, 061601\relax
\mciteBstWouldAddEndPuncttrue
\mciteSetBstMidEndSepPunct{\mcitedefaultmidpunct}
{\mcitedefaultendpunct}{\mcitedefaultseppunct}\relax
\EndOfBibitem
\bibitem[Park and Kim(2008)]{park2008_insects}
K.~J. Park and H.-Y. Kim, \emph{Journal of Fluid Mechanics}, 2008,
  \textbf{610}, 381--390\relax
\mciteBstWouldAddEndPuncttrue
\mciteSetBstMidEndSepPunct{\mcitedefaultmidpunct}
{\mcitedefaultendpunct}{\mcitedefaultseppunct}\relax
\EndOfBibitem
\bibitem[Vella(2008)]{Vella2008}
D.~Vella, \emph{Langmuir}, 2008, \textbf{24}, 8701--8706\relax
\mciteBstWouldAddEndPuncttrue
\mciteSetBstMidEndSepPunct{\mcitedefaultmidpunct}
{\mcitedefaultendpunct}{\mcitedefaultseppunct}\relax
\EndOfBibitem
\bibitem[Bush and Hu(2006)]{bush2006}
J.~W. Bush and D.~L. Hu, \emph{Annual Review of Fluid Mechanics}, 2006,
  \textbf{38}, 339--369\relax
\mciteBstWouldAddEndPuncttrue
\mciteSetBstMidEndSepPunct{\mcitedefaultmidpunct}
{\mcitedefaultendpunct}{\mcitedefaultseppunct}\relax
\EndOfBibitem
\bibitem[Burton and Bush(2012)]{burton2012}
L.~J. Burton and J.~W.~M. Bush, \emph{Physics of Fluids}, 2012, \textbf{24},
  101701\relax
\mciteBstWouldAddEndPuncttrue
\mciteSetBstMidEndSepPunct{\mcitedefaultmidpunct}
{\mcitedefaultendpunct}{\mcitedefaultseppunct}\relax
\EndOfBibitem
\bibitem[Landau and Lifshitz(1970)]{landau_elasticity}
L.~Landau and E.~Lifshitz, \emph{Theory of Elasticity}, Pergamon Press,
  1970\relax
\mciteBstWouldAddEndPuncttrue
\mciteSetBstMidEndSepPunct{\mcitedefaultmidpunct}
{\mcitedefaultendpunct}{\mcitedefaultseppunct}\relax
\EndOfBibitem
\bibitem[Hertz(1884)]{hertz1884}
H.~Hertz, \emph{Annalen der Physik}, 1884, \textbf{258}, 449--455\relax
\mciteBstWouldAddEndPuncttrue
\mciteSetBstMidEndSepPunct{\mcitedefaultmidpunct}
{\mcitedefaultendpunct}{\mcitedefaultseppunct}\relax
\EndOfBibitem
\bibitem[F\"oppl(1897)]{foppl}
A.~F\"oppl, \emph{Vorlesungen \"uber technische Mechanik}, B. G. Teubner,
  Leipzig, 1897\relax
\mciteBstWouldAddEndPuncttrue
\mciteSetBstMidEndSepPunct{\mcitedefaultmidpunct}
{\mcitedefaultendpunct}{\mcitedefaultseppunct}\relax
\EndOfBibitem
\bibitem[Timoshenko(1940)]{timoshenko}
S.~Timoshenko, \emph{Strength of Materials}, D. van Nostrand, New York,
  1940\relax
\mciteBstWouldAddEndPuncttrue
\mciteSetBstMidEndSepPunct{\mcitedefaultmidpunct}
{\mcitedefaultendpunct}{\mcitedefaultseppunct}\relax
\EndOfBibitem
\bibitem[Michael(1981)]{Michael1981}
D.~H. Michael, \emph{Annual Review of Fluid Mechanics}, 1981, \textbf{13},
  189--216\relax
\mciteBstWouldAddEndPuncttrue
\mciteSetBstMidEndSepPunct{\mcitedefaultmidpunct}
{\mcitedefaultendpunct}{\mcitedefaultseppunct}\relax
\EndOfBibitem
\bibitem[Rivetti and Neukirch(2012)]{Rivetti2012}
M.~Rivetti and S.~Neukirch, \emph{Proceedings of the Royal Society A:
  Mathematical, Physical and Engineering Science}, 2012, \textbf{468},
  1304--1324\relax
\mciteBstWouldAddEndPuncttrue
\mciteSetBstMidEndSepPunct{\mcitedefaultmidpunct}
{\mcitedefaultendpunct}{\mcitedefaultseppunct}\relax
\EndOfBibitem
\bibitem[de~Gennes(1985)]{gennes1985}
P.~G. de~Gennes, \emph{Reviews of Modern Physics}, 1985, \textbf{57},
  827--863\relax
\mciteBstWouldAddEndPuncttrue
\mciteSetBstMidEndSepPunct{\mcitedefaultmidpunct}
{\mcitedefaultendpunct}{\mcitedefaultseppunct}\relax
\EndOfBibitem
\bibitem[Diamant and Witten(2011)]{diamant2011}
H.~Diamant and T.~A. Witten, \emph{Physical Review Letters}, 2011,
  \textbf{107}, 164302--\relax
\mciteBstWouldAddEndPuncttrue
\mciteSetBstMidEndSepPunct{\mcitedefaultmidpunct}
{\mcitedefaultendpunct}{\mcitedefaultseppunct}\relax
\EndOfBibitem
\bibitem[Rivetti(2013)]{rivetti_cras}
M.~Rivetti, \emph{C. R. M{\'e}canique}, 2013, \textbf{341}, 333 -- 338\relax
\mciteBstWouldAddEndPuncttrue
\mciteSetBstMidEndSepPunct{\mcitedefaultmidpunct}
{\mcitedefaultendpunct}{\mcitedefaultseppunct}\relax
\EndOfBibitem
\bibitem[Lake and Thomas(1967)]{Lake1967}
G.~J. Lake and A.~G. Thomas, \emph{Proc. R. Soc. London A}, 1967, \textbf{300},
  108--119\relax
\mciteBstWouldAddEndPuncttrue
\mciteSetBstMidEndSepPunct{\mcitedefaultmidpunct}
{\mcitedefaultendpunct}{\mcitedefaultseppunct}\relax
\EndOfBibitem
\bibitem[Besson and Debr{\'e}geas(2007)]{Besson2007}
S.~Besson and G.~Debr{\'e}geas, \emph{The European Physical Journal E: Soft
  Matter and Biological Physics}, 2007, \textbf{24}, 109--117\relax
\mciteBstWouldAddEndPuncttrue
\mciteSetBstMidEndSepPunct{\mcitedefaultmidpunct}
{\mcitedefaultendpunct}{\mcitedefaultseppunct}\relax
\EndOfBibitem
\bibitem[Poivet \emph{et~al.}(2003)Poivet, Nallet, Gay, and Fabre]{Poivet2003}
S.~Poivet, F.~Nallet, C.~Gay and P.~Fabre, \emph{EPL (Europhysics Letters)},
  2003, \textbf{62}, 244\relax
\mciteBstWouldAddEndPuncttrue
\mciteSetBstMidEndSepPunct{\mcitedefaultmidpunct}
{\mcitedefaultendpunct}{\mcitedefaultseppunct}\relax
\EndOfBibitem
\end{mcitethebibliography}

\end{document}